\documentclass[namedreferences]{kluwer}
\input{epsf.sty}
\begin{document}
\begin{article}
% Now you can start to edit the file    
\begin{opening}
\title{Radial Gradients of Abundances}
\author{Mercedes \surname{Moll\'{a}}}
\author{Angeles I.\surname{D\'{\i}az}}
\institute{Dept. F\'{\i}sica Te\'{o}rica, 
Universidad Aut\'{o}noma de Madrid, 28049 Cantoblanco, Spain}
\author{F. \surname{Ferrini}}
\institute{INTAS, 58 Avenue des Arts, 1000 Bruxelles, Belgium}

\runningauthor{Moll\'{a}, D\'{\i}az \& Ferrini}
\runningtitle{Radial Gradients of Abundances}

\begin{abstract}
We have computed a set of multiphase chemical evolution models in
which the radial mass distribution of each theoretical galaxy is
calculated using the universal rotation curve from Persic, Salucci \&
Steel (1996). We obtain the chemical evolution for galaxies of
different masses and morphological types by changing the efficiencies
to form molecular clouds and stars according with these types.

We obtain the radial distribtions of diffuse and molecular gas
densities, the star formation rate and abundances for 15 elements for
each galaxy.

\end{abstract}
\end{opening}

\section{The generalization of the multiphase evolution model}

The multiphase chemical evolution model assumes a spherical
protogalaxy with a gas mass which collapses to fall onto the
equatorial plane forming the disk as a secondary structure.

The radial distribution of mass of each theoretical galaxy is
calculated through the Universal Rotation Curve from
\inlinecite{pss96}.  The infall rate of gas from a halo region to the
disk is inversely proportional to a collapse time scale which depends
on the total mass being shorter for the more massive galaxies.

The stars in the disks form in two steps: molecular clouds form from
the diffuse gas by a Schmidt law, and then cloud-cloud collisions
produce stars by a spontaneous process.  The efficiencies
$\epsilon_{\mu}$ and $\epsilon_{h}$ to form clouds or stars by these
two processes are assumed to depend on morphological or Hubble type T,
such as it was found in our previous investigations. For futher
details, see \inlinecite{mol01}.

\section{Results: The behaviour of oxygen abundances radial gradients}

Under these assumptions, we obtain the radial distributions of diffuse
and molecular gas densities, the star formation rate and abundances
for 15 elements for each galaxy.

\begin{figure}

\includegraphics{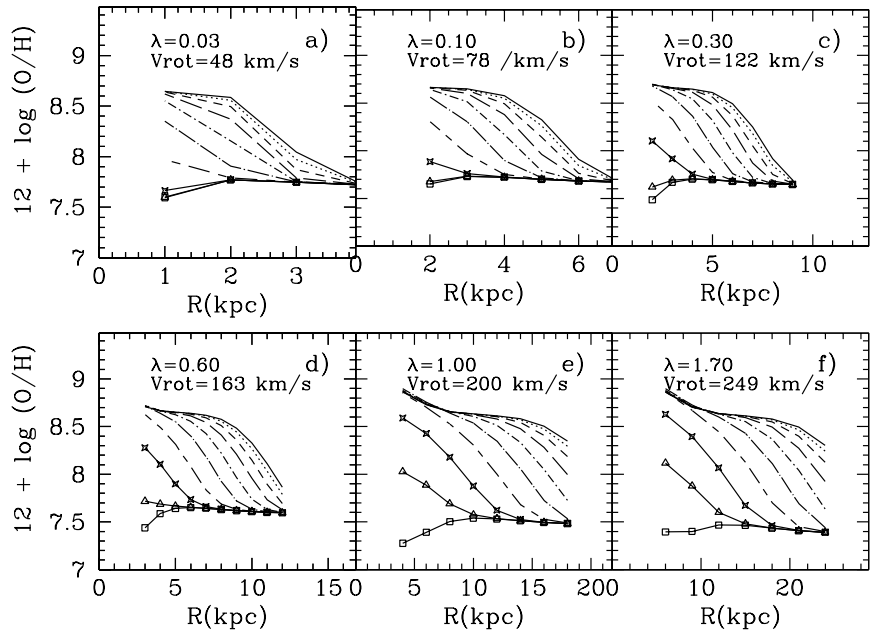}
\vspace*{10cm}
\caption{The oxygen abundance radial distributions for six different
values of galaxy mass. En each panel 10 different models are represented
corresponding to 10 Hubble T, with the most evolved ones at the top 
and the less ones at the bottom.}
\label{oh}
\end{figure}

The most important outcome concerns
the oxygen abundance: the radial gradient only appears for the
intermediate types ($7 < \rm{T} < 4$) at all galaxy masses, being
larger for the less massive galaxies. However, the latest ones (T $
\ge 8$) show a flat gradient with abundances $12 +\log{(O/H)}\sim$
7.5-8. This solves the apparent inconsistency of the largest gradients
appearing in late type spirals while some irregulars shows no gradient
at all. 

Results, which can see in Fig.~\ref{oh}, are in close
agreement with observations for HSB, LSB and dwarf
galaxies. Variations of the radial gradients of abundances with the
morphological type and with the total mass of the
galaxy follow the observed correlations.

\end{article}
\end{document}